\documentclass{ifacconf}

\usepackage[dvipdfmx]{graphicx}      
\usepackage{natbib}        
\usepackage{amsmath, amssymb}

\usepackage{color}
\newtheorem{prb}[thm]{Problem}
\usepackage{mathtools}

\usepackage{url}

\begin{document}
\begin{frontmatter}

\title{Sampled-Data State Observation over Lossy Networks under Round-Robin Scheduling } 


\author[First]{Toshihide Tadenuma} 
\author[First]{Masaki Ogura} 
\author[First]{Kenji Sugimoto}

\address[First]{Nara Institute of Science and Technology, 
Nara, Japan \\ (e-mail: tadenuma.toshihide.tk8@is.naist.jp, oguram@is.naist.jp, kenji@is.naist.jp).}

\begin{abstract}                
In this paper, we study the problem of continuous-time state observation over lossy communication networks. We consider the situation in which the samplers for measuring the output of the plant are spatially distributed and their communication with the observer is scheduled according to a round-robin scheduling protocol. We allow the observer gains to dynamically change in synchronization with the scheduling of communications. In this context, we propose a linear matrix inequality (LMI) framework to design the observer gains that ensure the asymptotic stability of the error dynamics in continuous time. We illustrate the effectiveness of the proposed methods by several numerical simulations.
\end{abstract}


\end{frontmatter}

\section{Introduction}

Networked control systems (NCSs) are feedback control systems in which the control loop is closed by shared communication channels. Although NCSs provide us with flexibilities in designing control systems, the use of shared communication channels can deteriorate the performance of the NCSs. For example, when using a shared communication channel, the control systems may no longer preserve stability due to unavoidable packet dropouts. For dealing with such problems in NCSs, several pieces of research have been done in the past decades~\citep{Hespanha2007a}. In this direction, we find a plethora of studies for the analysis and synthesis of NCSs under realistic communication protocols. For example, \cite{Nesic2004a} analyzed the input-output stability of NCSs for a large class of network scheduling protocols. \cite{Sinopoli2005} designed the optimal controller and estimator for NCSs under TCP- and UDP-like communication channels, respectively. \cite{Lin2015} studied the LQG control and Kalman filtering of NCSs in which packet acknowledgments from an actuator to an estimator is not available.

The round-robin scheduling provides a simple but yet effective communication protocol and is widely used in practice \citep{Behera2010,Joshi2015,Datta2015}. The round-robin protocol is the transmission rule in which each data is transmitted one by one in a fixed circular order. It reduces transmission rate and results in low data collision and packet dropouts.NCSs with round-robin protocol have been modeled by several approaches, such as time-delay approaches~\citep{Liu2015}, impulsive approaches and input-output approaches~\citep{Tabbara2008}. \cite{Donkers2011} studied stability of continuous-time NCSs with round-robin and try-once-discard protocol considering time-varing transmission intervals.
\cite{Xu2013} analyzed the stability of discrete-time NCSs with round-robin scheduling protocols and packet dropouts.\cite{Liu2015} analyzed input-to-state stability of NCSs with round-robin or try-once-discard protocol considering time-varing delays and transmission intervals.
\cite{Zou2016a} proposed a method for set-membership observation of discrete-time NCSs under round-robin and weighted try-once-discard protocols, respectively. However, there is scarce of design methodologies for the continuous-time NCSs under round-robin scheduling. 

In this paper, we present a linear matrix inequalities (LMIs) approach to design a sampled-data observer for continuous-time linear time-invariant systems under round-robin scheduling protocols and packet dropouts. Under the assumption that the number of successive packet dropouts is uniformly bounded, we present LMIs for finding the observer gains ensuring the asymptotic stability of the error dynamics in continuous time. To deal with packet dropouts, we allow the observer to dynamically change its observer gains depending on the time elapsed from the last measurement~\citep{Ogura2015f}. In order to derive the LMIs, we use a switched quadratic Lyapunov function for the estimation errors in discrete-time~\citep{Ding2009}.

This paper is organized as follows. After presenting the notation used in this paper, in Section~\ref{sec:problemFormulation} we state the sampled-data observation problem studied in this paper. In Section~\ref{sec:mainResults}, we present and prove the main result. We confirm the effectiveness of the main result in Section~\ref{sec:examples}.

\subsection*{Notations} For a real function~$f$, we let $f(t^-)$ denote the left-hand limit of $f$ at $t$. Let $\mathbb{R}, \mathbb{R}^n$ and $\mathbb{R}^{n\times m}$ denote the set of real numbers, real $n$-dimensional vectors and $n\times m$ real matrices, respectively. Let $\mathbb{N}$ denote the set of nonnegative integers. The $n\times m$ zero matrix is denoted by $0_{n\times m}$, or by~$0$ when $n$ and $m$ are clear from the context. Let $I_n$ denote the $n\times n$ identity matrix. The transpose of a matrix~$M$ is denoted by $M^\top$. If $M$ is square, then we define $\textbf{He}(M)=M+M^\top$. If $M$ is positive definite (negative definite), then we write $M\succ0$ ($M\prec0$, respectively). Let $\lambda_\mathrm{max(min)}(M)$ denote the largest (smallest, respectively) eigenvalue of $M$. The symbol~$*$ is used to denote the symmetric blocks of partitioned symmetric matrices. 

We close this section by stating Finsler's Lemma:
\begin{lem}[Finsler's Lemma] \label{Finsler}
Let $\xi\in\mathbb{R}^n,\ \mathcal{P}=\mathcal{P}^\top\in\mathbb{R}^{n\times n}$, and $\mathcal{H}\in\mathbb{R}^{m\times n}$. Assume that $\mathrm{rank}(\mathcal{H})=r<n$. The following statements are equivalent:
\begin{enumerate}
\item If $\xi\ne0$ satisfies $\mathcal{H}\xi=0$, then $\xi^\top \mathcal{P} \xi<0$. 
\item There exists $\mathcal{X}\in \mathbb{R}^{n \times m}$ such that  $\mathcal{P}+\mathcal{X}\mathcal{H}+\mathcal{H}^\top\mathcal{X}^\top\prec 0$. 
\end{enumerate}
\end{lem}

\begin{figure}[tb]
\begin{center}
\includegraphics[width=1.0\linewidth]{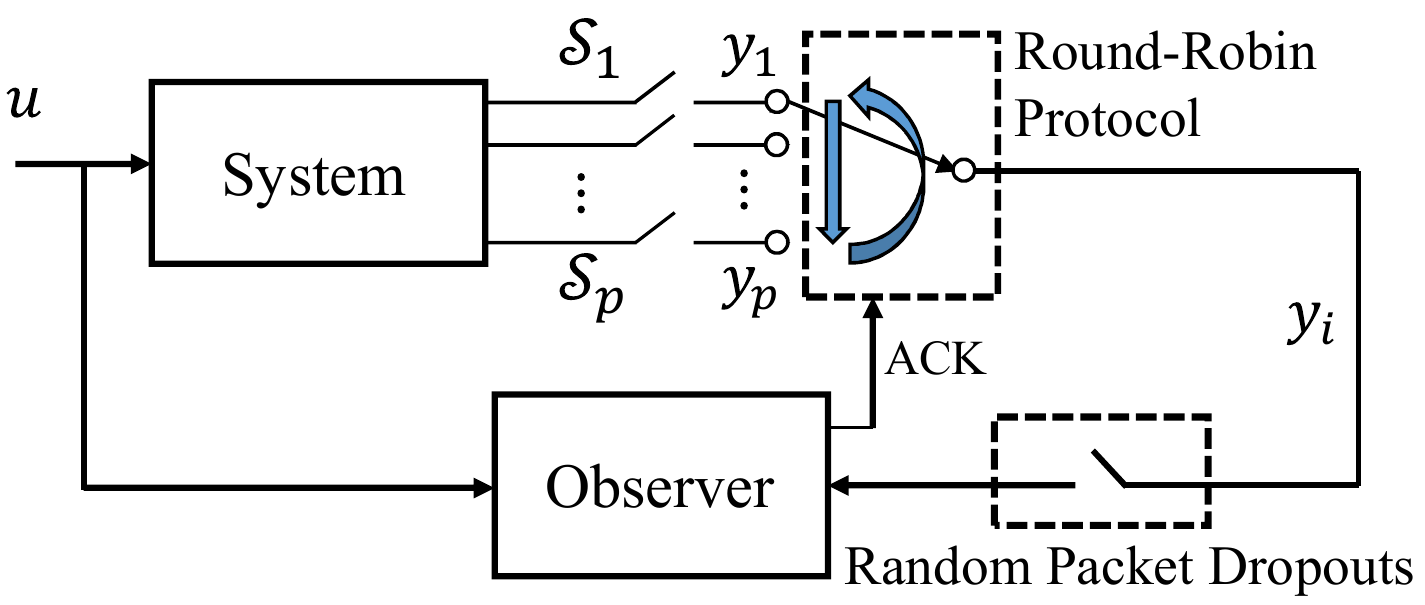}   
\caption{Networked control system under Round-Robin scheduling.} 
\label{fig:block}
\end{center}
\vspace{3mm}
\end{figure}

\section{Problem Formulation}\label{sec:problemFormulation}

In this section, we formulate the problem of sampled-data observations of linear time-invariant systems over networks with a round-robin scheduling and packet losses. 

\begin{figure}[tb]
\begin{center}
	\includegraphics[width=.87\linewidth]{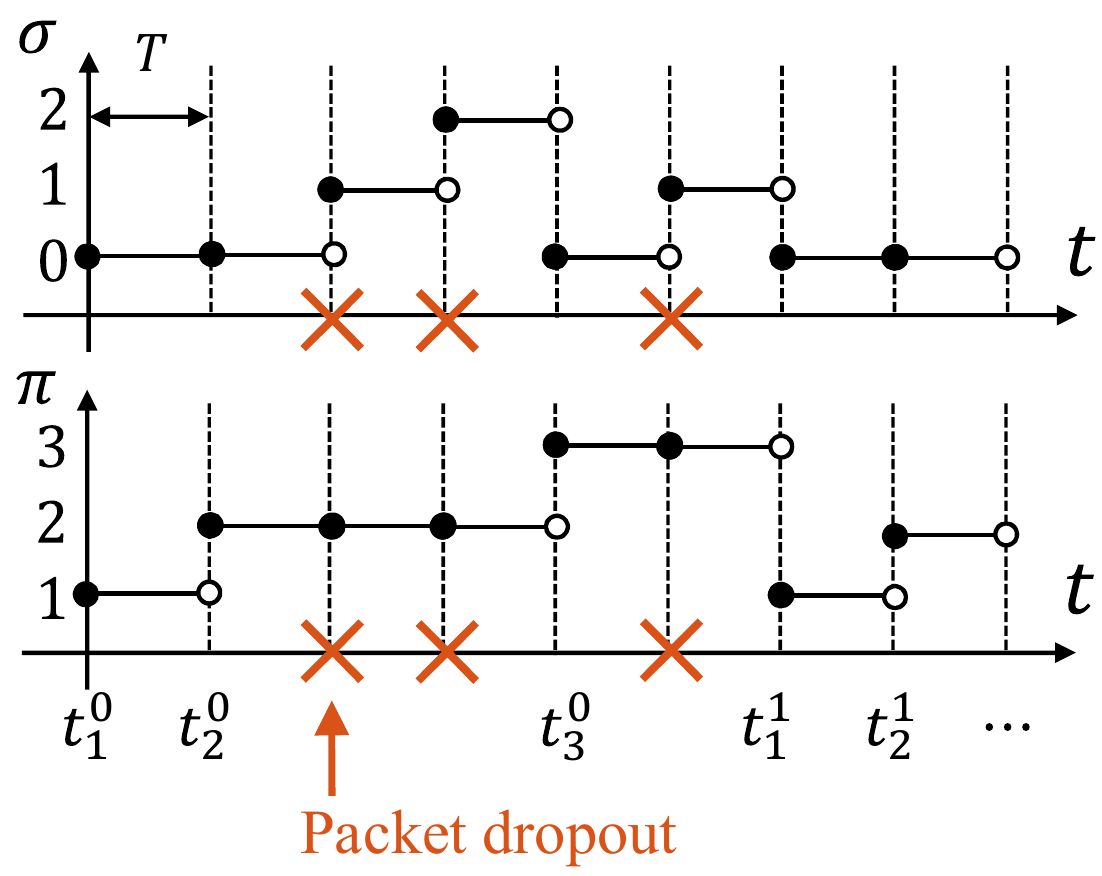}
	\caption{Dynamics of the counters $\sigma$ and $\pi$ for the case $p=3$. } 
	\label{fig:sigma}
\end{center}
\vspace{3mm}
\end{figure}

We consider the linear time-invariant system
\begin{equation} \label{LTI}
\Sigma: \begin{cases}
\begin{aligned}
\frac{dx}{dt} &= Ax+Bu,  \\
y & =Cx, 
\end{aligned}
\end{cases}
\end{equation}
where $x(t)\in \mathbb{R}^n$, $u(t)\in \mathbb{R}^m$ and~$y(t) \in \mathbb{R}^p$ are the state, the input and the output of the system and $A$, $B$ and~$C$ are real matrices having appropriate dimensions. 
As described in Fig.~\ref{fig:block}, for each $i = 1, \dotsc, p$, we assume that an ideal sampler~$\mathcal S_i$ periodically measures the $i$th component~$y_i$ of the output signal~$y$. We assume that the samplers share the same period~$T>0$ and are spatially distributed. Therefore, their communications with the observer have to be appropriately scheduled to avoid possible conflicts. We specifically follow the formulations in \cite{Zou2016,Li2017} and consider the scenario in which a round-robin protocol schedules the communications. The scheduling of the communications between the observer and samplers is described as follows (see Fig.~\ref{fig:sigma} for a schematic representation):
\begin{enumerate}
\item Let $\pi(0) = 1$ and $\sigma(0) = 0$. 
\item At each time $t=kT$ ($k=0, 1, \dotsc$), the $\pi(kT)$th sampler $\mathcal S_{\pi(kT)}$ sends the sample $y_{\pi(kT)}(kT)$ to the observer, while the other samplers take no action. 
\begin{itemize}
\item If the observer receives the sample, then we update the values of the counters $\pi$ and $\sigma$ as 
\begin{equation*}
\pi(kT) = \begin{cases}
1,& \mbox{if $\pi(kT^-) = p$}, 
\\
\pi(kT^-)+1,& \mbox{otherwise}, 
\end{cases}
\end{equation*}
and
\begin{equation*}
\sigma(kT) = 0. 
\end{equation*}
The values of $\pi$ and $\sigma$ are kept constant until the next sampling instant. 
\item If the observer does not receive the sample due to a packet dropout, then the sensor is acknowledged and will re-send a sample at the next time instant. In this case, the counters are updated as 
\begin{equation*}
\pi(kT) = \pi(kT^-)
\end{equation*}
and 
\begin{equation*}
\sigma(kT) = \sigma(kT^-)+1. 
\end{equation*}
The values of $\pi$ and $\sigma$ are kept constant until the next sampling instant.
\end{itemize}
\item Step (2) is repeated for each $k$ . 
\end{enumerate}

For each $i=1$, \dots, $p$, we let $t^0_i$, $t^1_i$, $t^2_i$, \dots denote the times at which the observer receives a measurement from the $i$th sampler. In this paper, we assume $t_1^0 = 0$ for simplicity of presentation, although the results presented in this paper hold true without this assumption. Then, by the cyclicity in the round-robin scheduling, the observer receives measurements from the samplers at the following time instants: 
\begin{equation}\label{eq:mewsuretimes}
0=t_1^0 < t_2^0 < \cdots < t_p^0 <\cdots < t_1^{h} < t_2^{h} < \cdots < t_p^{h} < \cdots. 
\end{equation}
Therefore, the information received by the observer is represented by the sequence 
\begin{equation*}
\begin{aligned} 
&y_1(t_1^0),\ y_2(t_2^0),\ \dotsc,\ y_p(t_p^0),\dotsc\\
&\quad y_1(t_1^{h}),\ y_2(t_2^{h}),\ \dotsc,\ y_p(t_p^{h}), \dotsc. 
\end{aligned}
\end{equation*}

We then describe the dynamics of the observer to be designed. For each $t\geq 0$, let $\tau(t)$ denote the most recent time at which a sample is received by the observer before time $t$. Mathematically, $\tau(t)$ is defined by
\begin{equation*}
\tau(t) = t_i^{h}
\end{equation*}
if 
\begin{equation*}
t \in \begin{cases}
[t_i^{h}, t_{i+1}^{h}), & \mbox{if $i\neq p$}, 
\\
[t_p^{h}, t_1^{{h}+1}),& \mbox{otherwise}. 
\end{cases}
\end{equation*}
Then, at each time $t$, the observer has the following pieces of information: (i) the counters $\pi(t)$ and $\sigma(t)$, (ii) the input $u(t)$ and (iii) the most recent sample $y_{\pi(t)}(\tau(t))$. We assume that the observer updates its estimate of the state variable, denoted by $\hat x(t) \in \mathbb{R}^n$, by the following differential equation
\begin{equation} \label{eq:observer}
\frac{d\hat x}{dt}=A\hat{x}(t)+Bu(t)+L_{\pi(t)}^{\sigma(t)}\bigl(y_{\pi(t)}(\tau(t))-c_{\pi(t)}\hat{x}(\tau(t))\bigr), 
\end{equation}
where $c_i$ denotes the $i$th row of the matrix~$C$ and
\begin{equation}\label{eq:Lid}
L_i^d \in \mathbb{R}^n \ (i \in \{1, \dotsc, p\},\, d \in \mathbb{N} )
\end{equation}
are the observer gains to be designed.

We can now state the problem studied in this paper: 

\begin{prb}\label{prb:}
For all ${h}\in\mathbb{N}$ and $i \in \{1, \dotsc, p\}$, let $d_i^h$ denote the number of successive packet dropouts during the time interval~$[t_i^{h}, t_{i+1}^{h}]$ (~$[t_p^{h}, t_{1}^{h+1}]$ in the case of ~$i=p$). Given the linear time-invariant system~$\Sigma$ and the sampling period~$T$, find the observer gains in \eqref{eq:observer}, \eqref{eq:Lid} such that 
\begin{equation}\label{eq:deferroraspstability}
\lim_{t\to\infty}(x(t)-\hat x(t)) = 0
\end{equation}
for all initial states $x(0), \hat x(0) \in \mathbb{R}^n$ as well as any pattern of successive packet dropouts $\{d_i^h\}_{i=1, \dotsc, p,\,h\in \mathbb{N}}$.  
\end{prb}

We remark that Problem~\ref{prb:} is not solvable if we allow packet dropouts to occur successively for infinitely many times. In order to avoid this pathological situation, we place the following assumption on the uniform boundedness on the number of successive packet dropouts, which is commonly adopted in the literature \citep[see, e.g., ][]{Zhang2008,Wang2010d}:
\begin{assum}
There exists a nonnegative integer $\bar d$ such that the numbers of successive packet dropouts are at most~$\bar d$. 
\end{assum}

\section{Observer Design}\label{sec:mainResults}

In this section, we state and prove the main result of this paper. We specifically show that we can reduce Problem~\ref{prb:} to a set of linear matrix inequalities, which can be efficiently solved.  

\subsection{Main Result}

The following theorem enables us to solve Problem~\ref{prb:} by solving a set of linear matrix inequalities and is the main result of this paper: 

\begin{thm}\label{theorem1num}
Let $\lambda \ne0 \in \mathbb{R}$. 
For each $d = 0, \dotsc, \bar d$ and $i = 1, \dotsc, p$, define the matrices 
\begin{equation}\label{eq:theMatrices}
	\begin{aligned}
\mathcal{A}_{d}&=e^{A(1+d)T},\\
\Gamma&=\int^{T}_{0}e^{A\tau}\,d\tau,\\
\mathcal{T}_{d}&=\Gamma
\begin{pmatrix}
\mathcal{A}_{d-1} & \mathcal{A}_{d-2} & \dotsc & \mathcal{A}_0 & I_{n}
\end{pmatrix} \in \mathbb{R}^{n\times n(1+d)}, \\
\mathcal{L}_i^d&=
\begin{pmatrix}
{L_i^0}^\top & {L_i^1}^\top & \dotsc & {L_i^d}^\top
\end{pmatrix}^\top \in \mathbb{R}^{n(1+d)}.
\end{aligned}
\end{equation}
Assume that, for all $i=1$, \dots, $p$ and $d, d' = 0$, \dots, $\bar d$, the $n\times n$ matrices~$P_i^d = (P_i^d)^\top\succ0$ and $X_i^d$ as well as the vectors $G_i^d\in \mathbb{R}^n$ satisfy the following LMIs:
\begin{align}\label{eq:mainLMI}
\begin{pmatrix}
-P_i^d+\textbf{He}(X_i^d\mathcal{A}_{d}-G_i^dc_i) & *  \\
-{X_i^d}^\top + \lambda(X_i^d\mathcal{A}_{d}-G_i^dc_i) & ~~P_{i+1}^{d'}-\lambda\textbf{He}(X_i^d)
\end{pmatrix}&\prec 0,\\
\mathrm{for}~~~i=1, \dots, p-1~~\mathrm{and} \nonumber\\
\begin{pmatrix}
-P_p^d+\textbf{He}(X_p^d\mathcal{A}_{h}-G_p^dc_p) & *  \\
-{X_p^d}^\top + \lambda(X_p^d\mathcal{A}_{d}-G_p^dc_p) & ~~P_1^{d'}-\lambda\textbf{He}(X_p^d)
\end{pmatrix} &\prec 0.   \nonumber \\
\label{theorem1}
\end{align}
Then, the observer gains defined by 
\begin{equation}\label{gain}
\begin{aligned}
L_i^0
&=
(X_i^0\Gamma)^{-1}G_i^0, 
\\
L_i^d&=(X_i^d\Gamma)^{-1}G_i^d- 
\begin{pmatrix}
\mathcal{A}_{d-1} & \mathcal{A}_{d-2} & \dotsc & \mathcal{A}_0
\end{pmatrix}\mathcal{L}_i^{d-1} 
\end{aligned}
\end{equation}
solve Problem~\ref{prb:}. 
\end{thm}

\begin{rem}
The invertibility of the matrix $X_i^d$ in \eqref{gain} is guaranteed by the lower-right block of the LMIs~\eqref{eq:mainLMI} and~\eqref{theorem1}. 
\end{rem}

\subsection{Proof}

In this subsection, we give the proof of Theorem~\ref{theorem1num}. We start from the following lemma on the dynamics of the estimation errors at the measurement times~\eqref{eq:mewsuretimes}:  

\begin{lem}\label{lem:}
Define $$\varepsilon = x-\hat{x}.$$ If $i \neq p$, then
\begin{equation}\label{eq:forLemma}
\varepsilon(t_{i+1}^{h})=(\mathcal{A}_{d_i^{h}}-\mathcal{T}_{d_i^{h}}\mathcal{L}_i^{d_i^{h}}c_i)\varepsilon(t_i^{h}). 
\end{equation}
If $i = p$, then 
\begin{equation} \label{error2}
\varepsilon(t_1^{{h}+1})=(\mathcal{A}_{d_p^{h}}-\mathcal{T}_{d_p^{h}}\mathcal{L}_p^{d_p^{h}}c_p)\varepsilon(t_p^{h}). 
\end{equation}
\end{lem}

\begin{pf}
We first consider the case $i \neq p$. Assume that $t_i^{h}\leq t<t_{i+1}^{h}$. From \eqref{LTI} and \eqref{eq:observer}, we obtain 
\begin{equation*} 
\dot{\varepsilon}(t)
=
A\varepsilon(t)-L_i^{\sigma(t)}c_i\varepsilon(t_i^{h}).  
\end{equation*}
This equation implies that 
\begin{equation*} 
\varepsilon(t)=e^{A(t-t_i^{h})} \varepsilon(t_i^{h})-\int_{t_i^{h}}^{t}e^{A(t-\tau)}L_i^{\sigma(\tau)}c_i\varepsilon(t_i^{h})\,d\tau.
\end{equation*}
Therefore, 
\begin{equation}  \label{conerror2}
\varepsilon(t_{i+1}^{h})=e^{A(1+d_i^{h})T} \varepsilon(t_i^{h})-\int_{t_i^{h}}^{t_{i+1}^{h}}e^{A(t_{i+1}^h-\tau)}L_i^{\sigma(\tau)}c_i\varepsilon(t_i^{h})\,d\tau,  
\end{equation}
where we have used $t_{i+1}^{h}-t_i^{h} = (1+d_i^{h})T$. The integral on the right hand side of equation~\eqref{conerror2} is rewritten as
\begin{equation*}
\begin{aligned}
&\int_{t_i^{h}}^{t_{i+1}^{h}}e^{A(t_{i+1}^{h}-\tau)}L_i^{\sigma(\tau)}\,d\tau c_i\varepsilon(t_i^{h}) \nonumber \\
=&\sum_{d=0}^{d_i^{h}}\int_{t_i^{h}+dT}^{t_i^{h}+(1+d)T}e^{A(t_{i+1}^{h}-\tau)}\,d\tau L_i^d c_i\varepsilon(t_i^{h}) \nonumber\\
=&\sum_{d=0}^{d_i^{h}}\int_{0}^{T}e^{A(t_{i+1}^{h}-t_i^{h}-dT-\tau)}\,d\tau L_i^d c_i\varepsilon(t_i^{h}) \nonumber\\
=&\sum_{d=0}^{d_i^{h}}\int_{0}^{T}e^{A((1+d_i^{h})T-dT-\tau)}\,d\tau L_i^d c_i\varepsilon(t_i^{h}). 
\end{aligned}
\end{equation*}
This equation and \eqref{conerror2} prove equation~\eqref{eq:forLemma} by \eqref{eq:theMatrices}. Equation~\eqref{error2} can be proved in the same manner and, therefore, the proof is omitted. 
\end{pf}

By using Lemma~\ref{lem:}, we can show that the observer gains given by Theorem~\ref{theorem1num} guarantee the convergence of the estimation error at the measurement time instants in~\eqref{eq:mewsuretimes}:

\begin{prop}
Let $\lambda\in \mathbb{R}\backslash \{0\}$ be arbitrary. Assume that the $n\times n$ matrices $P_i^d = (P_i^d)^\top \succ 0$ and $X_i^d$ as well as the vectors $G_i^d\in \mathbb{R}^n$ satisfy the LMIs in~\eqref{eq:mainLMI}. Define the observer gains by \eqref{gain}. Then, for all $i=1, \dotsc, p$, we have
\begin{equation}\label{eq:errorto0:disc}
\lim_{{h}\to\infty} \varepsilon(t_i^{h}) = 0
\end{equation}
for all initial states $x(0)$,\,$\hat x(0)$ and the numbers of successive packet dropouts $\{d_i^h\}_{i=1, \dotsc, p,\,h\in \mathbb{N}}$.  
\end{prop}

\begin{pf}
From \eqref{gain}, we obtain $G_i^0=X_i^0\Gamma L_i^0=X_i^0\mathcal{T}_{0}\mathcal{L}_i^0$ and
\begin{equation*}
\begin{aligned}
G_i^d&=X_i^d\Gamma L_i^d+X_i^d\Gamma
\begin{pmatrix}
\mathcal{A}_{d-1} & \mathcal{A}_{d-2} & \dotsc & \mathcal{A}_0
\end{pmatrix}
\begin{pmatrix}
L_i^0 \\ L_i^1 \\ \vdots \\ L_i^{d-1}
\end{pmatrix} \\
&=X_i^d\Gamma
\begin{pmatrix}
\mathcal{A}_{d-1} & \mathcal{A}_{d-2} & \dotsc & \mathcal{A}_0 & I_n
\end{pmatrix}
\begin{pmatrix}
L_i^0 \\ L_i^1 \\ \vdots \\ L_i^{d-1} \\ L_i^d
\end{pmatrix} \\
&=X_i^d\mathcal{T}_{d}\mathcal{L}_i^d.
\end{aligned}
\end{equation*}
Therefore, equation~\eqref{eq:mainLMI} shows 
\begin{equation}\label{proofLMIs1}
\begin{pmatrix}
-P_i^d+\textbf{He}(X_i^d(\mathcal{A}_{d}-\mathcal{T}_{d}\mathcal{L}_i^dc_i)) & *  \\
-{X_i^d}^\top+\lambda X_i^d(\mathcal{A}_{d}-\mathcal{T}_{d}\mathcal{L}_i^dc_i) & \ \ P_{i+1}^{d'}-\lambda\textbf{He}(X_i^d)
\end{pmatrix} \prec 0
\end{equation}
for all $i \neq p$ and $d, d' = 0$, \dots, $\bar d$. 
Define the matrices and vectors 
\begin{align}
\mathcal{P}_{i,d,j,d'}& = 
\begin{pmatrix}
-P_i^d & 0 \\
0 & P_j^{d'}
\end{pmatrix},\notag 
\\\notag 
\mathcal{X}_i^d&= 
\begin{pmatrix}
I_n  \\
\lambda I_n
\end{pmatrix}X_i^d,
\\ \notag 
\mathcal{H}_i^d&= 
\begin{pmatrix}
\mathcal{A}_d-\mathcal{T}_d\mathcal{L}_i^dC_i  & \ \ \ -I_n
\end{pmatrix},
\\ \label{eq:def:xi}
\xi(t_i^{h})&=
\begin{pmatrix}
\varepsilon(t_i^{h}) \\ \varepsilon(t_{i+1}^{h})
\end{pmatrix},\ \xi(t_p^{h})=
\begin{pmatrix}
\varepsilon(t_p^{h}) \\ \varepsilon(t_{1}^{{h}+1})
\end{pmatrix}. 
\end{align}
Then, the LMI~\eqref{proofLMIs1} is rewritten as
\begin{equation*}
\mathcal{P}_{i,d,i+1,d'}+\mathcal{X}_i^d\mathcal{H}_i^d+{\mathcal{H}_i^d}^\top{\mathcal{X}_i^d}^\top\prec 0. 
\end{equation*}
Also, equations \eqref{eq:forLemma} and \eqref{error2} show that
\begin{align*}
\mathcal{H}_i^{d_i^{h}}\xi(t_i^{h})=0. 
\end{align*}
Therefore, by Lemma~\ref{Finsler} we obtain 
\begin{equation*}
\xi(t_i^{h})^\top \mathcal{P}_{i,d,i+1,d'}\xi(t_i^{h})<0~~\mathrm{for}~~\xi(t_i^{h})\ne0. 
\end{equation*} 
This inequality shows that
\begin{equation}\label{eq:ineq1}
\varepsilon(t_{i+1}^{h})^\top P_{i+1}^{d_{i+1}^{h}}\varepsilon(t_{i+1}^{h})-\varepsilon(t_i^{h})^\top P_{i}^{d_i^{h}}\varepsilon(t_i^{h})<0 
\end{equation}
by the definition of $\xi$ in \eqref{eq:def:xi}. 
In a similar manner, we can show that 
\begin{equation}\label{eq:ineq2}
\varepsilon(t_1^{{h}+1})^\top P_1^{d_{1}^{{h}+1}}\varepsilon(t_1^{{h}+1})-\varepsilon(t_p^{h})^\top P_p^
{d_p^{h}}\varepsilon(t_p^{h})<0.
\end{equation}

Since the matrix $P_i^d$ is positive definite, inequalities~\eqref{eq:ineq1} and \eqref{eq:ineq2} show $$\lim_{{h}\to\infty} \varepsilon(t_{i}^{h})^\top P_{i}^{d_{i}^{h}}\varepsilon(t_{i}^{h}) = 0~~\mathrm{for}~~i=1,\dots,p.$$ Furthermore, the following inequalities hold.
$$\lambda_{\mathrm{min}}(P_i^{d_i^h}) \|z\|^2 \leq z^\top P_i^{d_i^h}z \leq \lambda_{\mathrm{max}}(P_i^{d_i^h}) \|z\|^2,  ~z\in\mathbb{R}^n.$$
Therefore, for any pattern of successive packet dropouts $\{d_i^{h}\}_{i=1, \dotsc, p, h\geq 0}$, we have  $\lim_{{h}\rightarrow\infty}\|\varepsilon(t_i^{h})\|=0$ for all $i=1$, \dots, $p$. This completes the proof of the proposition.
\end{pf}

We can now prove our main result:

\def\Elproofname{Proof of Theorem~\ref{theorem1num}.}
\begin{pf}
For all $i=1$, \dots, $p$ and $t\in[0,\bar{d}T]$, define the matrix 
\begin{equation*}
M_i(t)=e^{At}-\int_{0}^{t}e^{A(t-\tau)}L_i^{\sigma(\tau)}\,d\tau c_i.  \end{equation*}
Let 
\begin{equation*}
\alpha=\max_{0\leq t\leq\bar{d}T,\ i\in\{1,2,\dotsc,p\}} \sqrt{\lambda_{\mathrm{max}}(M_i(t)^\top M_i(t))}.
\end{equation*}
Notice that $\alpha$ is finite because $M_i$ is a continuous function. 
Then, from \eqref{conerror2}, if $t_i^{h}\leq t<t_{i+1}^{h}$, then we have 
\begin{equation*}
\|\varepsilon(t)\|=\|M_i(t-t_i^{h})\varepsilon(t_i^{h})\|\leq \alpha\|\varepsilon(t_i^{h})\|. 
\end{equation*}
Therefore, equation~\eqref{eq:errorto0:disc} implies $\lim_{t\to\infty}\varepsilon(t) = 0$. Therefore, equation~\eqref{eq:deferroraspstability} holds true as desired.
\end{pf}

\subsection{Concentrated Samplers} 
In this section, in order to clarify features of our proposed design, we briefly consider for comparison the case {\it without} round-robin scheduling, namely, the case where the samplers $\mathcal S_1$, \dots, $\mathcal S_p$ are spatially concentrated and, therefore, can simultaneously communicate with the observer in a synchronized fashion. In this situation, we consider the following (standard) communication protocol:

\begin{enumerate}
\item[($1'$)] Let $\sigma(0) = 0$. 
\item[($2'$)] At each time $t=kT$ ($k=0, 1, \dotsc$), the samplers $\mathcal S_1$, \dots, $\mathcal S_p$ simultaneously send the samples $y_{1}(kT)$, \dots, $y_{p}(kT)$ to the observer. 
\begin{itemize}
\item If the observer receives the samples, then we update the values of the counter $\sigma$ as 
$\sigma(kT) = 0. $
The value of $\sigma$ is kept constant until the next sampling instant. 
\item If the observer does not receive the samples due to a packet dropout, then the sensor is acknowledged and will re-send a sample at the next time instant. In this case, the counter is updated as 
$\sigma(kT) = \sigma(kT^-)+1$. 
The value of $\sigma$ is kept constant until the next sampling instant.
\end{itemize}
\item[($3'$)] Step ($2'$) is repeated for each $k$ . 
\end{enumerate}

As in the case of spatially distributed samplers described in Section~\ref{sec:problemFormulation}, we let $t^0$, $t^1$, $t^2$, \dots denote the times at which the observer receives measurements from the samplers. We assume $t^0 = 0$ for simplicity of presentation. 
For each $t\geq 0$, let $\tau(t)$ denote the most recent time at which samples are received by the observer before time $t$.
Then, we consider the observer 
\begin{equation}
\frac{d}{dt}\hat{x}(t) = A\hat{x}(t)+Bu(t)+L_{\sigma(t)}(y(\tau(t))-C\hat{x}(\tau(t))), \label{corob}
\end{equation}
where 
\begin{equation}\label{eq:Lid:concentraited}
L_d \in \mathbb{R}^{n\times p} \ (d \in \{0, \dotsc, \bar{d}\})
\end{equation} 
are the gains to be designed. Our objective in this section is to design the observer gains \eqref{eq:Lid:concentraited} that achieves the asymptotic stability~\eqref{eq:deferroraspstability} of the error for any pattern of successive packet dropouts $\{d^h\}_{h\in \mathbb{N}}$.  

By following the same argument as in the proof of Theorem~\ref{theorem1num}, we can prove the following corollary for designing the observer gains~\eqref{eq:Lid:concentraited}. The proof is omitted.

\begin{cor}
Let $\lambda\in \mathbb{R}\backslash \{0\} $ be arbitrary. Assume that the $n\times n$ matrices $P_d = (P_d)^\top \succ 0$ and $X_d$ as well as the matrices $G_d\in \mathbb{R}^{n\times p}$ satisfy the following LMIs
\begin{equation*}
\begin{pmatrix}
-P_d+\textbf{He}(X_d\mathcal{A}_{d}-G_dC) & *  \\
-{X_d}^\top+\lambda(X_d\mathcal{A}_{d}-G_dC) & \ \ P_{d'}-\lambda\textbf{He}(X_d)
\end{pmatrix} \prec 0
\end{equation*}
for all $d,d' = 0$, \dots, $\bar{d}$. 
Define
\begin{equation*}
\begin{aligned}
L_0&=(X_0\Gamma)^{-1}G_0,  \\
L_d&=(X_d\Gamma)^{-1}G_d- 
\begin{pmatrix}
\mathcal{A}_{d-1} & \mathcal{A}_{d-2} & \dotsc & \mathcal{A}_0
\end{pmatrix} \mathcal{L}_{d-1}. 
\end{aligned}
\end{equation*}
Then, we have \eqref{eq:deferroraspstability}
for all initial states $x(0), \hat x(0) \in \mathbb{R}^n$ as well as any pattern of successive packet dropouts $\{d^h\}_{h\in \mathbb{N}}$.  
\end{cor}

\section{Numerical Examples} \label{sec:examples}

In this section, we numerically illustrate Theorem~\ref{theorem1num}. Let 
\begin{align*}
A &= 
\begin{pmatrix}
0.05&~-0.59&~1.04&~2.14\\
0.57&~-0.26&~-0.26&~-0.62\\
-1.05&~1.36&~-0.62&~1.51\\
-1.48&~-1.01&~-0.35&~0.09
\end{pmatrix},\\
C &= 
\begin{pmatrix}
c_1\\
c_2
\end{pmatrix}=
\begin{pmatrix}
1&0&0&0\\
0&1&0&0
\end{pmatrix}. 
\end{align*}
The matrix $B$ is omitted in this simulation because we adopt $u(t)=0$ for simplicity. The system~$\Sigma$ is unstable because $A$ has the eigenvalues  $0.1965\pm2.1188i$ having positive real parts. 

\begin{figure}[tb]
	\begin{center}
		\includegraphics[width=1\linewidth]{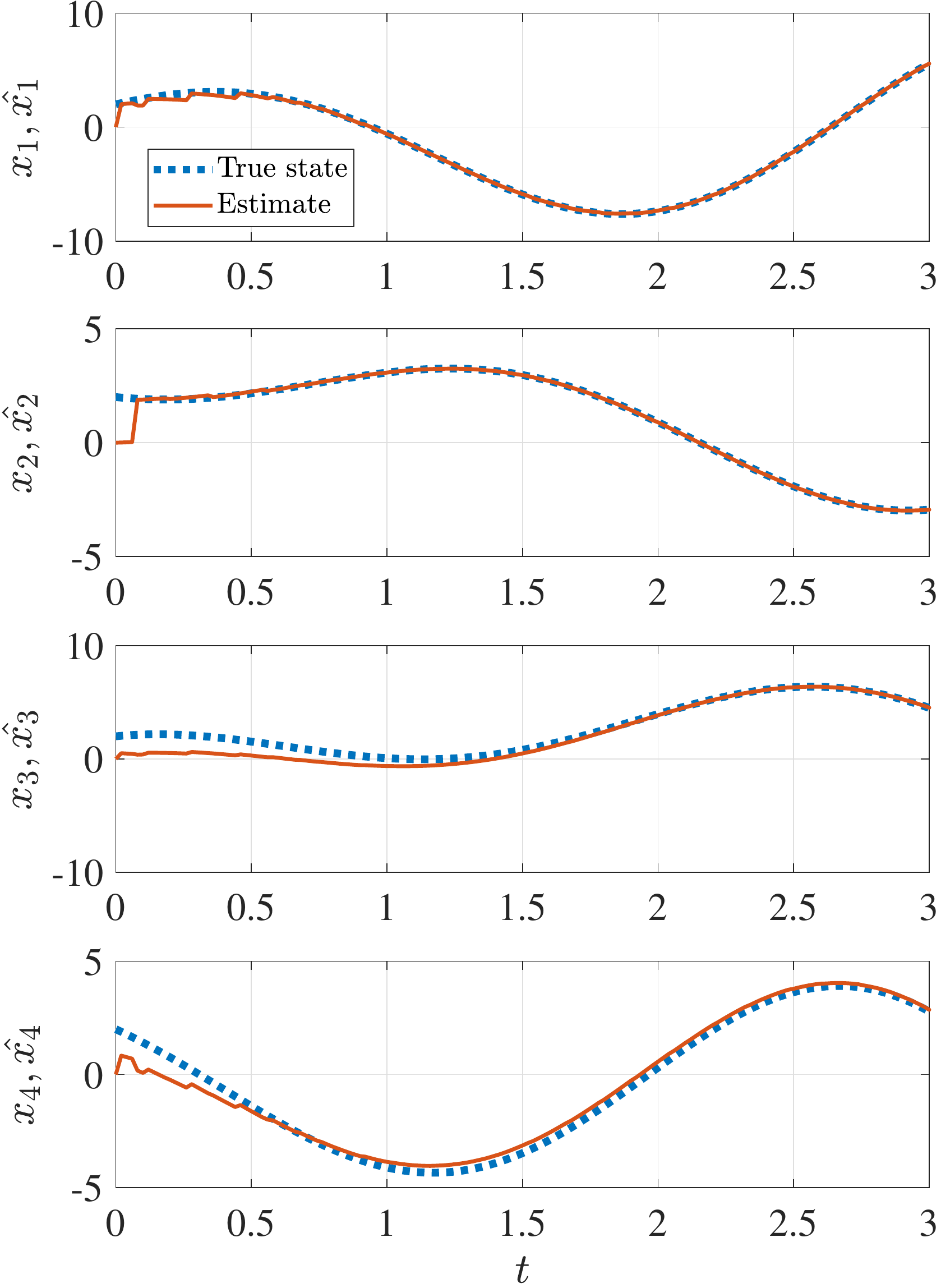}    
		\caption{True state~$x$ and its estimate $\hat x$. Dotted lines: $x$. Solid lines: $\hat x$.	}
		\label{fig:x1234}
	\end{center}
\vspace{3mm}
\end{figure}

\begin{figure}[tb]
	\begin{center}
		\hspace{3mm}
		\includegraphics[width=.95\linewidth]{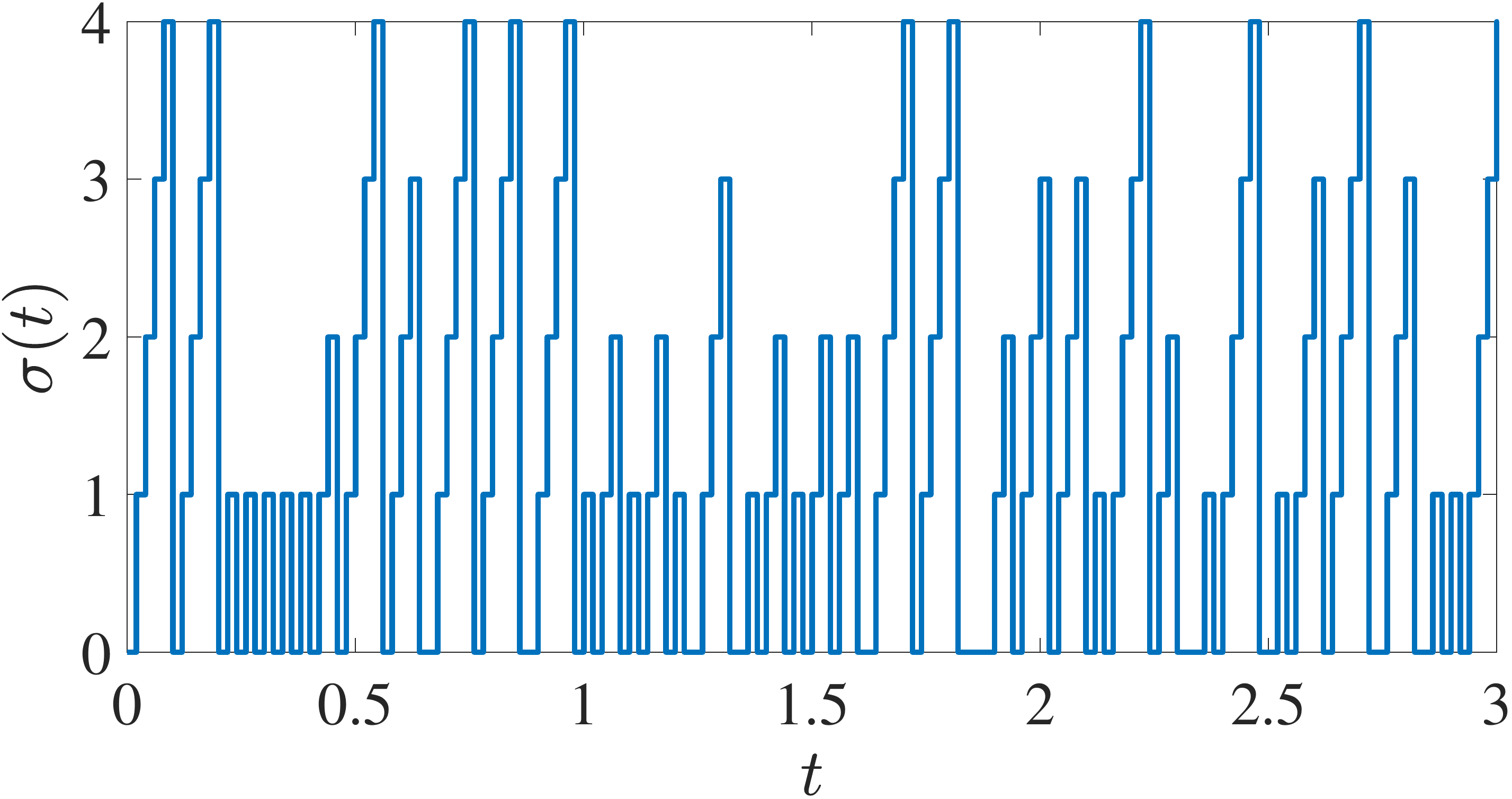}    
		\caption{The counter~$\sigma(t)$. The numbers of successive packet dropouts are drawn from a uniform distribution on the set~$\{0,1,\dotsc,4\}$ }
		\label{fig:sigmaplot}
	\end{center}
\vspace{3mm}
\end{figure}

Let $T=0.02$ and $\bar{d}=4$. We design the observer gains by solving the LMIs in \eqref{eq:mainLMI} with the parameter~$\lambda = 20$, using MATLAB~R2017b, YALMIP  and Mosek~8.
In Fig.~\ref{fig:x1234}, we show the trajectories of the state~$x$ and its estimate~$\hat x$ for the initial states~$x(0)=(2~~2~~2~~2)^\top$ and $\hat{x}(0)=0$ as well as the packet dropouts illustrated in Fig.~\ref{fig:sigmaplot}. In this simulation, the number of packet dropout $d_i^h$ is determined at the sampling time $t_i^h$ by uniform distribution on the set $\{0,1,\dotsc,4\}$.  We see that our state observation is successful even under heavy packet dropouts as shown in Fig.~\ref{fig:sigmaplot}. The drawback seen in Fig.~\ref{fig:x1234} is its slow convergence but it can be improved by optimizing decay rate or $\gamma$ performance \citep{Ding2009}.



\begin{figure}[tb]
	\begin{center}
		\hspace{1.5mm}\includegraphics[width=.98\linewidth]{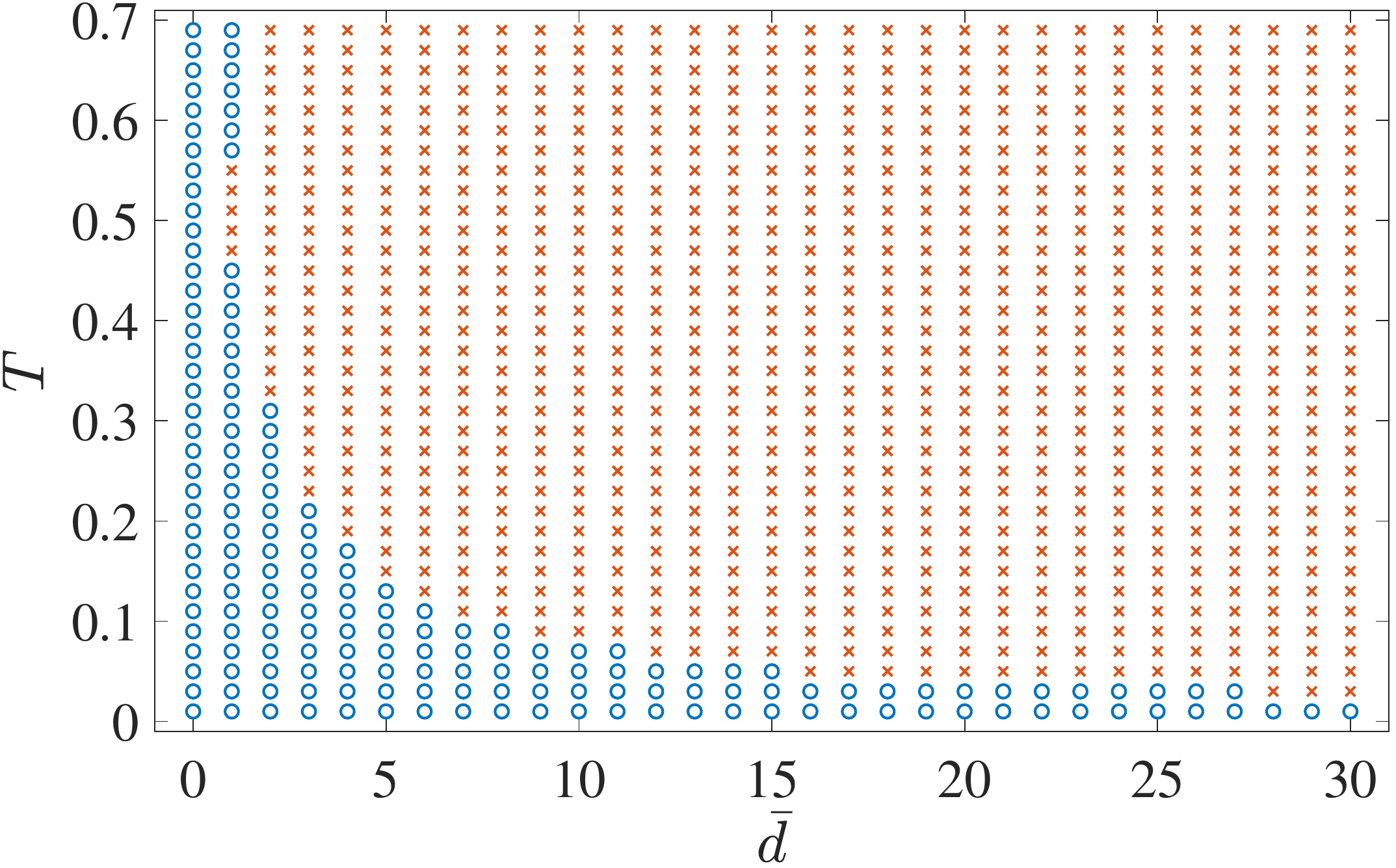}    
		\caption{Solvabilities of LMIs \eqref{eq:mainLMI} for various values of $\bar d$ and $T$. Blue circles: solvable, Red cross marks: not solvable.}
		\label{fig:Tdplot}
	\end{center}
\vspace{3mm}
\end{figure}

Then, in order to examine the relationship between the length of the sampling period ($T$) and the maximum number of successive packet dropouts ($\bar d$), we examine the solvability of the LMIs~\eqref{eq:mainLMI} for various values of $T$ and $\bar d$. The results are shown in Fig.~\ref{fig:Tdplot}. We confirm that the larger $\bar d$ the shorter sampling period~$T$ is required for successful state observations. We also remark an interesting phenomenon where, in the case of $\bar d=1$, the smaller $T$ does not necessarily improve our ability to design an observer. This phenomenon indicates the possible conservativeness of our formulation, which is left as an open problem.

\section{Conclusion}

In this paper, we have studied the problem of sampled-data observation for continuous-time linear-time invariant systems over lossy communication networks. We have specifically considered the situation in which the samplers for measuring the output of the plant are spatially distributed and their communications with the observer are scheduled according to a round-robin protocol. By using a switched quadratic Lyapunov function, we have presented LMIs for designing the switching gains of the observer. The effectiveness of the proposed methods has been illustrated by numerical simulations. 


\begin{thebibliography}{15}
\providecommand{\natexlab}[1]{#1}
\providecommand{\url}[1]{\texttt{#1}}
\providecommand{\urlprefix}{URL }
\expandafter\ifx\csname urlstyle\endcsname\relax
  \providecommand{\doi}[1]{doi:\discretionary{}{}{}#1}\else
  \providecommand{\doi}{doi:\discretionary{}{}{}\begingroup
  \urlstyle{rm}\Url}\fi

\bibitem[{Behera et~al.(2010)Behera, Mohanty, and Nayak}]{Behera2010}
Behera, H.S., Mohanty, R., and Nayak, D. (2010).
\newblock {A new proposed dynamic quantum with re-adjusted round robin
  scheduling algorithm and its performance analysis}.
\newblock \emph{International Journal of Computer Applications}, 5(5), 10--15.

\bibitem[{Datta(2015)}]{Datta2015}
Datta, L. (2015).
\newblock {Efficient round robin scheduling algorithm with dynamic time slice}.
\newblock \emph{International Journal of Education and Management Engineering},
  5(2), 10--19.

\bibitem[{Ding and Yang(2009)}]{Ding2009}
Ding, D.W., and Yang, G.H. (2009).
\newblock {Static output feedback control for discrete-time switched linear
  systems under arbitrary switching}.
\newblock \emph{2009 American Control Conference}, 2385--2390.

\bibitem[{Donkers et~al.(2011)Donkers, Hetel, Heemels, Wouw, and Steinbuch}]{Donkers2011}
Donkers, M.C.F., Hetel, L., Heemels, W.P.M.H., van de Wouw, N., and Steinbuch, M. (2011).
\newblock {Stability analysis of networked control systems using a switched systems approach}.
\newblock \emph{IEEE Transactions on Automatic Control}, 56(9), 2101--2115.

\bibitem[{Hespanha et~al.(2007)Hespanha, Naghshtabrizi, and Xu}]{Hespanha2007a}
Hespanha, J.P., Naghshtabrizi, P., and Xu, Y. (2007).
\newblock {A survey of recent results in networked control systems}.
\newblock \emph{Proceedings of the IEEE}, 95(1), 138--162.

\bibitem[{Joshi and Tyagi(2015)}]{Joshi2015}
Joshi, R., and Tyagi, S.B. (2015).
\newblock {Smart optimized round robin (SORR) CPU scheduling algorithm}.
\newblock \emph{International Journal of Advanced Research in Computer Science
	and Software Engineering}, 5(7), 568--574.

\bibitem[{Li et~al.(2017)Li, Lu, Xu, Peng, and Rao}]{Li2017}
Li, J.Y., Lu, R., Xu, Y., Peng, H., and Rao, H.X. (2017).
\newblock {Distributed state estimation for periodic systems with sensor
  nonlinearities and successive packet dropouts}.
\newblock \emph{Neurocomputing}, 237, 50--58.

\bibitem[{Lin et~al.(2015)Lin, Su, Shi, Lu, and Wu}]{Lin2015}
Lin, H., Su, H., Shi, P., Lu, R., and Wu, Z.G. (2015).
\newblock {LQG control for networked control systems over packet drop links
  without packet acknowledgment}.
\newblock \emph{Journal of the Franklin Institute}, 352(11), 5042--5060.

\bibitem[{Liu et~al.(2015)Liu, Fridman, and Hetel}]{Liu2015}
Liu, K., Fridman, E., and Hetel, L. (2015).
\newblock {Networked control systems in the presence of scheduling protocols and communication delays}.
\newblock \emph{SIAM Journal on Control and Optimization}, 53(4), 1768--1788.

\bibitem[{Nesic and Teel(2004)}]{Nesic2004a}
Nesic, D., and Teel, A.R. (2004).
\newblock {Input-Output stability properties of networked control systems}.
\newblock \emph{IEEE Transactions on Automatic Control}, 49(10), 1650--1667.

\bibitem[{Ogura et~al.(2018)Ogura, Cetinkaya, Hayakawa, and
  Preciado}]{Ogura2015f}
Ogura, M., Cetinkaya, A., Hayakawa, T., and Preciado, V.M. (2018).
\newblock {State feedback control of Markov jump linear systems with
  hidden-Markov mode observation}.
\newblock \emph{Automatica}, 89, 65--72.

\bibitem[{Sinopoli et~al.(2005)Sinopoli, Schenato, Franceschetti, Poolla, and
  Sastry}]{Sinopoli2005}
Sinopoli, B., Schenato, L., Franceschetti, M., Poolla, K., and Sastry, S.
  (2005).
\newblock {An LQG optimal linear controller for control systems with packet losses}.
\newblock \emph{44th IEEE Conference on Decision and Control},
  458--463.
  
 \bibitem[{Tabbara and Nesic(2008)}]{Tabbara2008}
 Tabbara, M., and Nesic, D. (2008).
 \newblock {Input–output stability of networked control systems with stochastic protocols and channels}.
 \newblock \emph{IEEE Transactions on Automatic Control}, 53(5), 1160--1175.

\bibitem[{Wang et~al.(2010)Wang, Han, and Yu}]{Wang2010d}
Wang, Y.L., Han, Q.L., and Yu, X. (2010).
\newblock {Packet dropout separation-based networked control systems
  quantitative synthesis}.
\newblock \emph{49th IEEE Conference on Decision and Control},
  5875--5880.

\bibitem[{Xu et~al.(2013)Xu, Su, Pan, Wu, and Xu}]{Xu2013}
Xu, Y., Su, H., Pan, Y.J., Wu, Z.G., and Xu, W. (2013).
\newblock {Stability analysis of networked control systems with round-robin
  scheduling and packet dropouts}.
\newblock \emph{Journal of the Franklin Institute}, 350(8), 2013--2027.

\bibitem[{Zhang and Yu(2008)}]{Zhang2008}
Zhang, W.A., and Yu, L. (2008).
\newblock {Modelling and control of networked control systems with both
  network-induced delay and packet-dropout}.
\newblock \emph{Automatica}, 44, 3206--3210.

\bibitem[{Zou et~al.(2016{\natexlab{a}})Zou, Wang and Gao}]{Zou2016}
Zou, L., Wang, Z., and Gao, H. (2016{\natexlab{a}}).
\newblock {Observer-based $\mathrm{H}_\infty$ control of networked systems with stochastic
  communication protocol: The finite-horizon case}.
\newblock \emph{Automatica}, 63, 366--373.

\bibitem[{Zou et~al.(2016{\natexlab{b}})Zou, Wang and Gao}]{Zou2016a}
Zou, L., Wang, Z., and Gao, H. (2016{\natexlab{b}}).
\newblock {Set-membership filtering for time-varying systems with mixed
  time-delays under Round-Robin and Weighted Try-Once-Discard protocols}.
\newblock \emph{Automatica}, 74, 341--348.

\end{thebibliography}

\end{document}